\documentclass{article}
\usepackage{graphicx} 

\usepackage[margin=1in]{geometry}
\usepackage{pdflscape}
\usepackage{titlesec}
\usepackage[LGR,T1]{fontenc}
\usepackage{hyperref}
\hypersetup{
    colorlinks,
    linkcolor={red},
    citecolor={orange},
    urlcolor={blue!80!black}
}

\makeatletter
\def\bbordermatrix#1{\begingroup \m@th
  \@tempdima 4.75\p@
  \setbox\z@\vbox{%
    \def\cr{\crcr\noalign{\kern2\p@\global\let\cr\endline}}%
    \ialign{$##$\hfil\kern2\p@\kern\@tempdima&\thinspace\hfil$##$\hfil
      &&\quad\hfil$##$\hfil\crcr
      \omit\strut\hfil\crcr\noalign{\kern-\baselineskip}%
      #1\crcr\omit\strut\cr}}%
  \setbox\tw@\vbox{\unvcopy\z@\global\setbox\@ne\lastbox}%
  \setbox\tw@\hbox{\unhbox\@ne\unskip\global\setbox\@ne\lastbox}%
  \setbox\tw@\hbox{$\kern\wd\@ne\kern-\@tempdima\left[\kern-\wd\@ne
    \global\setbox\@ne\vbox{\box\@ne\kern2\p@}%
    \vcenter{\kern-\ht\@ne\unvbox\z@\kern-\baselineskip}\,\right]$}%
  \null\;\vbox{\kern\ht\@ne\box\tw@}\endgroup}
\makeatother

\usepackage{graphicx}
\usepackage{subcaption}
\usepackage{tikz}
\usepackage{pgfplots}
\pgfplotsset{compat=1.18} 
\usetikzlibrary{angles,
                graphs,
                graphs.standard, 
                quantikz2,
                decorations.pathmorphing,
                patterns, 
                patterns.meta,
                shadows}
\tikzset{snake it/.style={decorate, decoration=snake}}

\usepackage{physics}
\usepackage{amsmath,amsfonts, amsthm}
\usepackage{pifont}

\usepackage{csvsimple}
\usepackage{listings}
\usepackage{minted}

\usepackage{paralist}

\usepackage{booktabs}
\usepackage{makecell}

\usepackage{multirow}
\usepackage{mdframed}
    \newmdtheoremenv{defn}{Definition}
    \newmdtheoremenv{thrm}{Theorem}
    \newmdtheoremenv{thrm*}{Theorem}
    \newmdtheoremenv{hlo}{Overview}
    \newmdtheoremenv{lmma}{Lemma}
    \newmdtheoremenv{prop}{Proposition}
    \newmdtheoremenv{stm}{Statement}
    \newmdtheoremenv{corollary}{Corollary}
    \newmdtheoremenv{prob}{Problem}
    \newmdtheoremenv{rem}{Remark}

\usepackage{algorithm}
\usepackage{algpseudocode}   

\DeclareMathAlphabet{\mathgtt}{LGR}{cmtt}{m}{n}

\usepackage[toc,page]{appendix}
\usepackage[affil-it]{authblk}

\usepackage{biblatex}
\title{\Large{\textbf{Expressive and Scalable Quantum Fusion for Multimodal Learning}}}
\author[1,5]{\normalsize Tuyen Nguyen\thanks{\href{mailto:tuyen.q.nguyen@student.uts.edu.au}{tuyen.q.nguyen@student.uts.edu.au}}}
\author[2]{\normalsize Trong Nghia Hoang}
\author[3]{\normalsize Phi Le Nguyen}
\author[4]{\normalsize Hai L. Vu}
\author[5]{\normalsize Truong Cong Thang}
\affil[1]{\small University of Technology Sydney, NSW, Australia
}
\affil[2]{\small Washington State University, Washington, USA
}
\affil[3]{\small Hanoi University of Science and Technology, Hanoi, Vietnam
}
\affil[4]{\small Monash University, VIC, Australia
}
\affil[5]{\small The University of Aizu, Aizuwakamatsu, Japan
}
\date{}

\begin{document}

\maketitle
\begin{abstract}

The aim of this paper is to introduce a quantum fusion mechanism for multimodal learning and to establish its theoretical and empirical potential. The proposed method, called the Quantum Fusion Layer (QFL), replaces classical fusion schemes with a hybrid quantum–classical procedure that uses parameterized quantum circuits to learn entangled feature interactions without requiring exponential parameter growth. Supported by quantum signal processing principles, the quantum component efficiently represents high-order polynomial interactions across modalities with linear parameter scaling, and we provide a separation example between QFL and low-rank tensor-based methods that highlights potential quantum query advantages. In simulation, QFL consistently outperforms strong classical baselines on small but diverse multimodal tasks, with particularly marked improvements in high-modality regimes. These results suggest that QFL offers a fundamentally new and scalable approach to multimodal fusion that merits deeper exploration on larger systems.
\end{abstract}

\section{Introduction}
With recent state-of-the-art in self-driving cars \cite{xiao2020multimodal},  image and video understanding \cite{alayrac2022flamingo, sun2019videobert}, and text-to-image generation \cite{ramesh2021zero}, the success of multi-modal learning is an important further step in the development of a wide range of areas- robotics \cite{lee2019making, marge2022spoken}, human-computer interaction \cite{obrenovic2004modeling, sharma1998toward}, and healthcare \cite{cai2019survey, muhammad2021comprehensive}. 

However, multimodal learning faces the challenge of overfitting due to increased model capacity. Since modalities occupy distinct feature spaces and distributions, learning their joint representation often requires exponentially more parameters \cite{hou2019deep, zadeh2017tensor}. Wang \textit{et al.} \cite{wang2020makes} showed that simply doubling parameters by combining modalities can degrade performance without specialized fusion strategies. This underscores the importance of fusion mechanisms that not only align varying modality characteristics but also efficiently capture their high-dimensional interactions.

To address these challenges, recent research has focused on designing fusion techniques capable of modeling high-order cross-modal interactions. One prominent example is the tensor fusion network (TFN) introduced by \cite{zadeh2017tensor}, which explicitly models trilinear interactions through the full tensor product of modality-specific representations, leading to notable performance improvements. However, this expressiveness comes at the cost of exponential growth in dimensionality with the number of modalities, making such approaches impractical in high-modality regimes. To improve scalability, low-rank tensor approximations (i.e., low-rank multimodal fusion (LMF)) have been proposed \cite{liu2018efficient, hou2019deep}, which factorize the fusion tensor to reduce memory and computation costs. While effective in reducing complexity, these methods rely on an implicit separability assumption in the interaction space, which limits their ability to capture rich, entangled interactions that are inherently non-separable. More recently, graph-based fusion \cite{gao2020multi, peng2024learning} has emerged as a flexible alternative, representing modality features as graph nodes and learning interactions via message passing in Graph Neural Networks (GNNs) \cite{wu2022graph}. While GNNs can theoretically capture higher-order dependencies through stacked layers, they often rely on local aggregation and require deep architectures to model global interactions effectively. 

These limitations motivate the exploration of quantum architectures for multimodal fusion. The potential of quantum modeling arises from two fundamental characteristics of quantum systems: \textit{superposition} and \textit{entanglement}. Superposition enables the efficient encoding of global interactions without relying on locality assumptions, thus overcoming the inductive biases inherent in graph-based models. Entanglement, on the other hand, facilitates the modeling of complex, non-separable cross-modal dependencies without the restrictive separability assumptions required by low-rank tensor methods. Inspired by these observations, in this work, we introduce the Quantum Fusion Layer (QFL), a hybrid quantum-classical framework for multimodal learning that leverages recent advances in quantum signal processing and parameterized quantum circuits to efficiently capture arbitrary high-order polynomial interactions across modalities. The architecture, illustrated in Figure~\ref{fig:quantumfusionlayer}, embeds a quantum circuit into a standard multimodal pipeline, replacing the classical fusion layer. 

Our \textbf{main contributions} are threefold:  
\begin{enumerate}
    \item \textbf{Provable expressivity and complexity.} We prove that QFL can represent arbitrary multivariate polynomials over a high-dimensional torus with only linear growth in the number of parameters, providing (to our knowledge) the first rigorous expressivity theorem for quantum fusion. We also give a complete complexity analysis, with explicit bounds on the gate counts.  
    \item \textbf{Separation over Low-Rank Tensor Fusion (LMF).} As a first step toward demonstrating quantum advantages, we focus on tensor-based techniques, which are the natural baseline for formal comparison between quantum and classical models \cite{Rudolph_2023}. We prove a query-complexity separation showing that QFL achieves representational power unattainable by LMF under the same resource constraints. While graph-based fusion baselines such as GNNs are empirically strong, their lack of rigorous theoretical frameworks currently precludes formal separation results. Extending our analysis to such models, therefore, remains an important direction for future work. In this sense, our result represents the first separation theorem in multimodal fusion and lays the foundation for broader theoretical comparisons between quantum and classical fusion models.
    \item \textbf{Empirical Superiority.} We benchmark QFL against both LMF and GNN-based approaches across diverse tasks, ranging from low-modality settings (vision--language entailment) to high-modality scenarios (12-lead ECG, traffic forecasting with 207 sensors). QFL consistently outperforms classical baselines, with its strongest gains in high-modality regimes, where its theoretical advantages are most clearly reflected in practical performance.
\end{enumerate}


The remainder of this paper is organized as follows. In Section~\ref{sec: bg}, we provide background on polynomial interaction models for multimodal learning and introduce the framework of quantum signal processing that underpins our method. Section~\ref{sec: relatedwork} reviews related works in quantum multimodal learning. In Section~\ref{sec: approach}, we present the proposed Quantum Fusion Layer (QFL), detailing its architecture, design principles, and training procedure. Section~\ref{sec: theoretical} develops the theoretical analysis of QFL, including expressivity, parameter scaling, computational complexity, and a formal separation result over low-rank tensor methods. Section~\ref{sec: exp} reports extensive experiments across diverse multimodal benchmarks, highlighting both the scalability and empirical advantages of QFL. Finally, Section~\ref{sec: open_prob} concludes the paper and discusses open challenges for future research.
\section{Background}\label{sec: bg}
\subsection{High-order polynomial interaction for multimodal machine learning}
Multimodal fusion is one of the key steps in deciding the success of a multimodal machine learning model. Typically, there are three categories of multimodal fusion: early, late, and hybrid fusion \cite{baltruvsaitis2018multimodal}. While early fusion deals with the integration of inputs at the source data level \cite{d2015review}, late fusion aims to model each modality separately and then combine them at the decision level \cite{morvant2014majorityvotediverseclassifiers}. In this paper, we focus on the hybrid approaches, which is presented in Figure~\ref{fig:quantumfusionlayer}. Each modality will be processed by suitable unimodal encoders to extract a meaningful abstract representation. Then, they are combined and used to learn a joint representation. Finally, a decoder is applied to perform particular downstream tasks. The popular techniques for this kind of fusion are \textit{linear} or \textit{bilinear interactions}. Although these techniques are efficiently implemented and useful in several applications \cite{yu2017multi, mroueh2015deepmultimodallearningaudiovisual, wu2014exploring}, they may not be sufficient to capture the complicated intercorrelations among modalities. Several studies showed that higher-degree cross-modal interactions could be able to boost performance significantly \cite{zadeh2017tensor, hou2019deep}.

In particular, given a collection of features $\{\mathbf{z}^{(m)}\}_{m=1}^{M}$ of $M$ modalities such that $\mathbf{z}^{(m)} \in \mathbb{R}^{D}$, we aim to learn the joint representation that captures cross-modal interactions between data by exploiting the high-order moments. One common approach is representing the combined feature vectors $\mathbf{x}$ as the tensor product of unimodal feature vectors:
\begin{equation}
    \mathbf{x} := \begin{bmatrix}
        \mathbf{z}^{(1)} \\
        1
    \end{bmatrix} 
    \otimes 
    \begin{bmatrix}
        \mathbf{z}^{(2)} \\
        1
    \end{bmatrix}
    \otimes 
    \dots
    \otimes 
    \begin{bmatrix}
        \mathbf{z}^{(M)} \\
        1
    \end{bmatrix}
    \label{eqn: representation 2}
\end{equation}
The output vector $\mathbf{x} \in \mathbb{R}^{(D+1)^M}$ contains all possible polynomial monomials between unimodal embeddings. This approach has been shown to boost the performance significantly \cite{zadeh2017tensor, lin2015bilinear}. However, the approach limits the potential of fully representing multilinear feature intercorrelations by restricting the order of interactions (a fixed degree of $M$ in~\eqref{eqn: representation 2}). A more flexible framework is introduced in \cite{hou2019deep}, which allows us to present arbitrary high-order interactions between unimodal feature vectors. 

We denote a concatenated feature vector:
\begin{equation}
     \mathbf{x} := [1\ |\ {\mathbf{z}^{(1)}}\ |\ \mathbf{z}^{(2)}\ | ...|\ \mathbf{z}^{(M)}] \in \mathbb{R}^{MD+1} \label{eqn: framework}
\end{equation}
Then, an abritary $P$-degree polynomial feature vector is defined as the $P$-order tensor product of $ \mathbf{x}$:
\begin{equation}
    \mathbf{X}_{P} = \underbrace{ \mathbf{x} \otimes  \mathbf{x}\otimes \dots \otimes  \mathbf{x}}_{P-\text{times}}
    \label{eqn: representation 1}
\end{equation}
Again, $\mathbf{X}_P\in \mathbb{R}^{(MD+1)^P}$ could represent all possible polynomial monomials up to order $P$. Thus, the task of learning $P$-order polynomial interactions is learning the coefficients of the target function. Consider a weight tensor $\mathbf{W} \in \mathbb{R}^{(MD+1)^P\times H}$, the joint representation of the multimodal feature is represented as:
\begin{equation}
    \mathbf{f} = \mathbf{W}^T\mathbf{X}_P
\end{equation}
where $\mathbf{f}\in \mathbb{R}^{H}$ is a vector whose each element represents a $P$-degree polynomial of ($\mathbf{z}^{(1)}, \mathbf{z}^{(2)}, ..., \mathbf{z}^{(M)}$). Then, the problem is reducible to learning a linear model in a high-dimensional feature space of $(MD+1)^P$.

Observe that in either approach of~\eqref{eqn: representation 2} or~\eqref{eqn: representation 1}, the model requires a number of parameters ($\mathbf{W}$) that grows exponentially with the polynomial order to capture the full interactions between models. To tackle this issue, a common strategy is to utilize low-rank tensor networks to approximate $\mathbf{W}$, resulting in a polynomial scaling complexity \cite{zadeh2017tensor, hou2019deep, liu2018efficient}. However, on the functional level, low-rank tensor formats correspond to the separation of variables. Some common formats can be named as CP format \cite{jiang2021lowcpranktensorcompletionpractical}, functional tensor train format (FTT)\cite{dolgov2021functional}, and Tucker format \cite{bigoni2016spectral}. Each format has its own strengths and limitations \cite{grasedyck2013literaturesurveylowranktensor}, but all of them aim to approximate multivariate functions using a set of functions of each variable. For example, the CP format approximates a high-dimensional function $f(\{\mathbf{z}^{(m)}\}_{m=1}^{M})$ by expressing it as a sum of products of univariate functions: 
\begin{equation}
    f(\{\mathbf{z}^{(m)}\}_{m=1}^{M}) = \sum_{\alpha=1}^{R} g^{(1)}_{\alpha}(\mathbf{z}^{(1)}_1)g^{(2)}_{\alpha}(\mathbf{z}^{(1)}_2)\dots g^{(MD)}_{\alpha}(\mathbf{z}^{(M)}_D)
    \label{eqn: cp_format}
\end{equation}
That means if any function $f: [-1,1]^{n} \to \mathbb{R}$ can be written as a sum of separable functions as~\eqref{eqn: cp_format}, the function will be well-approximated with $\mathcal{O}(RMD)$ computational complexity \cite{beylkin2002numerical}. That allows tensor-network-based techniques to enjoy a polynomial scale complexity \cite{hou2019deep, liu2018efficient, zadeh2017tensor}. However, in practice, most functions are not typically separable; the value $R$ will scale exponentially to fully capture the complex relations. Here is where quantum entanglement becomes advantageous. An entangled system is one in which the quantum state cannot be expressed as a product of the states of its individual components \cite{strossner2024approximation}. Thus, the quantum settings might be able to bring us better approximation regions with appropriate entanglement controls. In this paper, we will propose a novel method utilizing the technique of quantum signal processing to efficiently implement the high-order polynomial interactions between multimodal data.



\subsection{Quantum Signal Processing}
Before going into further details of our algorithm, we will briefly introduce the framework of quantum signal processing and its powerful representation for any matrix-value polynomial \cite{martyn2021grand} on a complex unit circle. 

Quantum Signal Processing (QSP) or Quantum Signal Value Transform \cite{Gily_n_2019} currently stands as the most efficient technique for implementing polynomial functions of block-encoded matrices. The technique provides an abstract formalism that allows for the efficient implementation of numerous linear algebraic operations and transformations, which creates a unified language for quantum algorithm construction that includes three major algorithms of search, phase estimation, and Hamiltonian simulation \cite{martyn2021grand}. Given an input Hermitian matrix $\mathbf{A}$\footnote{In general, QSP works even $\mathbf{A}$ is not Hermitian.}, the core idea of QSP is to construct a polynomial $P(\mathbf{A})$ by assuming access to a unitary $\mathbf{U}_{\mathbf{A}}$ encoding $\mathbf{A}$ such that:
\begin{equation}
    \mathbf{U}_{\mathbf{A}} = \begin{bmatrix}
        \mathbf{A} & . \\
        . & .
    \end{bmatrix}
\end{equation}
Since the matrix $\mathbf{A} \in \mathbb{C}^{N \times N}$ has the eigenvalue decomposition of $\sum_i \sigma_i \ket{v_i}\bra{v_i}$, then a function $P: \mathbb{R} \rightarrow \mathbb{C}$ is defined as a matrix function: $ P(\mathbf{A}) := \sum_i P(\sigma_i) \ket{v_i}\bra{v_i}$. Then, the quantum signal processing algorithm will construct the quantum circuit parameterized by a vector of rotation angles $\Vec{\boldsymbol{\phi}}$, returning the block-encoding of the matrix polynomial $P(\mathbf{A})$ in the sense that:
\begin{equation}
    \text{QSP}(\mathbf{U}_{\mathbf{A}}, \Vec{\boldsymbol{\phi}}) = \begin{bmatrix}
        P(\mathbf{A}) & . \\
        . & .
    \end{bmatrix}
\end{equation}
The strength of QSP lies in its representational power. Indeed, Mothlagh \textit{et al.} \cite{motlagh2024generalized} proved that any univariate $d$-degree polynomial $P$ satisfying $|P|^2 \leq 1$ on a unit circle will be represented using the QSP framework using $2d+3$ parameters. The condition of $|P|^2 \leq 1$ is a necessary restriction due to the unitary nature of quantum computation. However, the current powerful construction of QSP is limited to univariate polynomials. To extend it into multivariate cases, a typical solution uses the Linear Combination of Unitaries (LCU) technique \cite{childs2012hamiltonian}. The separate terms of a multivariate polynomial are combined via LCU, which will induce significant overhead that might be proportional to the number of terms. Several studies attempt to address this issue by introducing multivariate quantum signal processing (M-QSP) in certain situations by directly combining the terms of a multivariate polynomial using a circuit similar to those used in univariate quantum signal processing \cite{nemeth2023variants, rossi2022multivariable}. A major barrier of the study in M-QSP is that the set of achievable multivariate polynomial transformations is quite limited; moreover, no good characterization was known for them. Thus, whether the representational power in QSP could be lifted to the multivariate case is still an open question. Recently, Rossi \textit{et al.} \cite{rossi2022multivariable} rely on a conjecture that Fejer-Riesz theorem (FRT) is equivalent to QSP. Based on this conjecture, the authors lifted to MQSP via the multivariate FRT and established a characterization of MQSP similar to QSP. However, this conjecture is refuted by Nemeth \textit{et al.}\cite{nemeth2023variants}, they provided a counterexample showing that the conjecture holds for bivariate polynomials that have a degree at most 1 in one of the variables, but does not hold in general. Alternatively, Nemeth \textit{et al.} proposed a construction for homogeneous bivariate (commuting) quantum signal processing where the desired characterization is still valid. However, it is unclear how it can be extended to homogeneous multivariate (non-commuting) variants. Inspired by that, we proposed a new construction that supports inhomogeneous multivariate polynomials, which are either commuting or non-commuting. The construction is then applied in the case of multimodal learning. We show that with linear growth of parameters, our construction can generate high-order polynomial interactions between multimodal data. 
\section{Related Works}\label{sec: relatedwork}
Quantum approaches to multimodal learning have recently gained attention. Early work by Li \textit{et al.}~\cite{li2021quantum} used complex-valued neural networks to mimic quantum operations for fusing modalities like text and video. More recent hybrid models employ parameterized quantum circuits (PQCs) to encode and integrate multimodal features. For example, QMNN~\cite{zheng2024quantum} and QMLSC~\cite{li2025qmlsc} use variational circuits to entangle visual, textual, and auditory information for tasks like sentiment analysis. Other efforts include QMFND~\cite{qu2024qmfnd} for fake news detection using quantum CNNs, QMCL~\cite{chen2024quantum} for contrastive learning over EEG-image pairs, and Pokharel \textit{et al.}~\cite{pokharel2025quantum} for quantum federated learning. However, these works mostly target bimodal or trimodal fusion, lack architectural scalability, and are supported primarily by empirical results with limited theoretical grounding. Consequently, the conditions under which quantum models offer advantages in expressivity and scalability remain poorly understood.
\section{Quantum Fusion Layer}\label{sec: approach}

In this section, we propose and present the design of the Quantum Fusion Layer (QFL). Inspired by principles from quantum signal processing (QSP), QFL constructs polynomial transformations through a sequence of multimodal state encoding and parameterized unitary operations. Each modality is first processed through a uni-modal encoder to extract features $\{\mathbf{z}^{(m)}\}_{m=1}^{M}$, which are then encoded into quantum states. These states are transformed via a parameterized quantum circuit, whose trainable parameters model high-order interactions. The resulting quantum measurements are passed to a classical decoder for prediction. Like other hybrid models, QFL is trained end-to-end using classical gradient-based optimization. To provide a clear understanding of the Quantum Fusion Layer, we next detail the design and highlight the role of each component. 

\begin{figure*}
    \centering
    \includegraphics[width=0.95\linewidth]{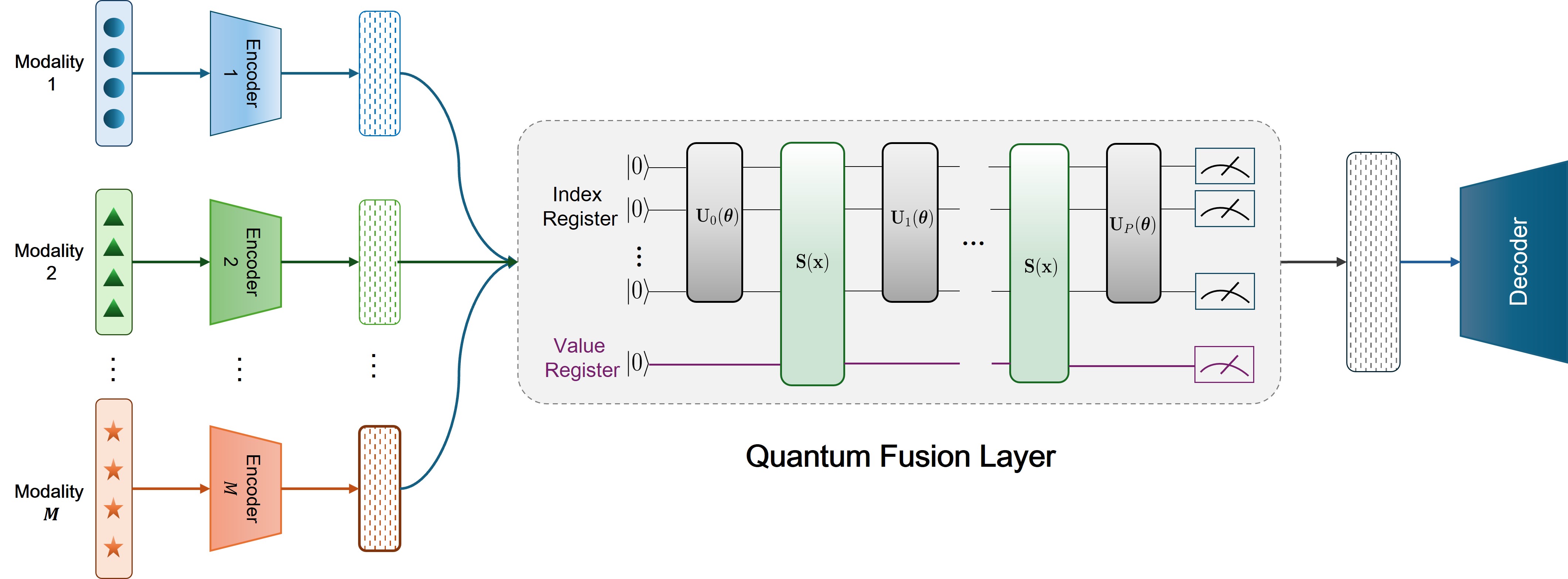}
    \caption[]{\textbf{Hybrid Architecture of Quantum Fusion Layer (QFL) for Multimodal Learning}. The QFL consists of three key components: (1) multimodal superposition state preparation \( \mathbf{S}(\mathbf{x}) \), 
    (2) parameterized quantum circuits \( \mathbf{U}(\boldsymbol{\theta}) \), and 
    (3) a measurement module. The sequence \(\mathbf{U}(\boldsymbol{\theta}), \mathbf{S}(\mathbf{x}) \) 
    is repeated \( P \) times to construct a degree-\( P \) multivariate polynomial over the input modalities.}
    \label{fig:quantumfusionlayer}
\end{figure*}
\paragraph{Multimodal superposition state preparation:}

Similar to the classical method, we first concatenate the multimodal feature vectors into:
\begin{equation}
    \mathbf{x} := [\ {\mathbf{z}^{(1)}}\ |\ \mathbf{z}^{(2)}\ | ...|\ \mathbf{z}^{(M)}\ ] \in \mathbb{R}^{MD}.
    \label{eqn: quantum_input}
\end{equation}
Given $\mathbf{x}$, we aim to design a unitary that performs the state preparation as follows: 
\begin{equation}
    \mathbf{S}(\mathbf{x}) := \ket{0}\bra{0}\otimes \mathbb{I} + \sum_{j=1}^{MD} \ket{j}\bra{j} \otimes \mathbf{R}_y(2\phi_j),
    \label{eqn: multimodal_unitary}
\end{equation}
where $\mathbb{I}$ is an identity matrix, $\phi_j =\arccos(\mathbf{x}_j)$, and $\mathbf{R}_y(\theta)$ is the rotation gate according to $y$-axis having the form of $e^{-i\sigma_y\theta/2 }$, where $\sigma_y$ is Pauli-Y matrix. The term of $\ket{0}\bra{0}\otimes \mathbb{I}$ will act as the constant term in~\eqref{eqn: framework} and allow us to represent all possible polynomial expansions rather than being restricted to homogeneous as in the design of \cite{nemeth2023variants}.  

In this work, we adopt the qubit-efficient encoding procedure~\cite{nguyen2022towards} to implement \( \mathbf{S}(\mathbf{x}) \). The method uses two arrays: an \textit{index register} and a \textit{value register}, which respectively encode the positions and values of the input tensor. For a given input \( \mathbf{x} \)~\eqref{eqn: quantum_input}, each dimension is assigned a separate index register, allowing the quantum encoding to preserve the \textit{geometric structure} of the data by associating positions with dedicated qubits. The value register is encoded using \textit{angle embedding} via a single-qubit rotation gate \( \mathbf{R}_y \)~\cite{Schuld_2021}. Letting \( n = \lfloor\log(MD+1)\rfloor \), the unitary \( \mathbf{S}(\mathbf{x}) \) acts on a {register} of \( n+1 \) qubits. The detailed procedure for implementing \( \mathbf{S}(\mathbf{x}) \) is outlined below:
\begin{itemize}    
    \item Uniform superposition of qubits in index register with Hadamard gates:
        \begin{equation}
            \mathbf{U}_H = H^{\otimes n} \otimes \mathbb{I}. 
        \end{equation}
    \item Encoding values: for each position $i \neq 0$, we apply a controlled rotation gate $\mathbf{R}_i$ by the angle $\phi_i = \arccos(\mathbf{x}_i)$:
        \begin{equation}
            \mathbf{R}_i = \left (\sum_{j=0, j\neq i}^{2^n}\ket{j}\bra{j} \otimes \mathbb{I}\right) +  \ket{i}\bra{i}\otimes \mathbf{R}_y(2\phi_i).
        \end{equation}
        Thus, we produce $\mathbf{S}(\mathbf{x})$ by applying $\mathbf{U}_H$ followed by $\prod_{i=1}^{2^n} \mathbf{R}_i$:
        \begin{equation}
            \mathbf{R}(\mathbf{x}) = \left(\prod_{i=1}^{2^n} \mathbf{R}_i\right)\mathbf{U}_H.
        \end{equation}
\end{itemize}
This technique generally requires a quantum circuit with $\mathcal{O}(MD)$ gates acting on $\lfloor\log(MD+1)\rfloor +1$ qubits. 

\paragraph{Parameterized Quantum Circuits:}

A variety of parameterized quantum circuit (PQC) architectures have been proposed, each tailored to specific application domains~\cite{romero2018strategies, kivlichan2018quantum, Schatzki_2024}. Among them, one of the most widely adopted designs is the hardware-efficient ansatz, introduced by Kandala \textit{et al.}~\cite{kandala2017hardware}, which prioritizes shallow depth and hardware feasibility. More recent works have also emphasized the importance of incorporating inductive biases into PQC design to enhance trainability and mitigate optimization challenges~\cite{Schatzki_2024, West_2024}.  

In general, the choice of circuit \( \mathbf{U}(\boldsymbol{\theta}) \) remains flexible and highly application-dependent. Within our proposed architecture, we apply \( \mathbf{U}(\boldsymbol{\theta}) \) exclusively to the index register, as shown in Figure~\ref{fig:quantumfusionlayer}. This design choice is inspired by quantum signal processing (QSP), where parameterized unitary operations act on an invariant subspace defined by the index register, enabling efficient implementation of structured polynomial transformations. By restricting parametrization to this register, QFL inherits both the scalability benefits of QSP circuits and the flexibility of PQCs for representing complex, high-order interactions across modalities.  

At the same time, QFL is not immune to the critical challenges that affect quantum neural networks more broadly. Issues such as barren plateaus and poor local minima~\cite{mcclean2018barren, Anschuetz_2022} may still arise in training, particularly as system size grows. However, several recent PQC families have been explicitly designed to mitigate these challenges, including problem-inspired ansätze~\cite{peruzzo2014variational}, locality-preserving designs such as the quantum convolutional neural network (QCNN)~\cite{cong2019quantum}, block-structured circuits like tensor-product or layered hardware-efficient ansätze with tailored initialization strategies~\cite{grant2019initialization, skolik2021layerwise}, or symmetry-imposed architecture~\cite{Schatzki_2024}.  

A key advantage of QFL is that it imposes no restriction on the specific form of \( \mathbf{U}(\boldsymbol{\theta}) \). Thus, one could seamlessly integrate these plateau-free PQC architectures within the QFL framework, combining the theoretical expressivity of quantum fusion with the practical benefits of more trainable circuit designs.

\paragraph{Measurement:}      
Measurement in quantum systems involves extracting classical information from quantum states for downstream tasks. In quantum machine learning, this is often done using a fixed computational basis~\cite{havlivcek2019supervised, schuld2020circuit}, which, while simple, may underutilize the representational richness of quantum states. Recent work~\cite{gao2022enhancing} shows that using diverse measurement bases can better reveal latent structure. Motivated by this, we introduce a \textit{random basis measurement} strategy to flexibly probe quantum states from multiple directions.

We build on the standard Pauli operators \( \sigma_x, \sigma_y, \sigma_z \), which correspond to spin measurements along the axes of the Bloch sphere. To move beyond fixed axes, we define randomized measurements in the planes spanned by Pauli pairs:
\[
\begin{aligned}
    \sigma_{xz} &= \cos\theta\, \sigma_x + \sin\theta\, \sigma_z, \\
    \sigma_{xy} &= \cos\phi\, \sigma_x + \sin\phi\, \sigma_y, \\
    \sigma_{yz} &= \cos\psi\, \sigma_y + \sin\psi\, \sigma_z,
\end{aligned}
\]
where \( \theta, \phi, \psi \in [0, 2\pi) \) are sampled randomly. Each operator \( \sigma_{ab} \) corresponds to a spin projection along a unit vector in the respective Bloch plane. By aggregating measurements over multiple random directions, we obtain an \textit{informationally complete} representation of the quantum state. The number of sampled directions determines the output dimensionality of the QFL.


\section{Theoretical Analysis} \label{sec: theoretical}
\subsection{Expressivity and Complexity}
In this subsection, we establish the theoretical results of QFL. The model is performed by repeating the application of the unitary $\mathbf{S}(\mathbf{x})$ and the parameterized transformation of $\mathbf{U}_p$\footnote{For notation simplicity, we ignore the set of parameters $\boldsymbol{\theta}_p$ in $\mathbf{U}_p$}:
\begin{equation}
    \mathbf{F}_{P}(\mathbf{x}) = \mathbf{U}_0.\mathbf{S}(\mathbf{x}).\mathbf{U}_1.\dots.\mathbf{U}_{P-1}.\mathbf{S}(\mathbf{x}).\mathbf{U}_{P}.  \label{eqn: mqsvt}
\end{equation}
where each of $\mathbf{U}_{p} \in SU(MD+1)$, which is the unitary group of degree $MD+1$. The key property of this proof is that the actions of $\mathbf{S}(\mathbf{x})$ and $\mathbf{U}_p$ perform transformations within the eigenspaces of the system. To see this, let us first derive the eigenvalues of $\mathbf{R}_y(2\theta)$ are $e^{\pm i \theta}$ corresponding to fixed eigenvectors what we denote as $\ket{v^{+}}$ and $\ket{v^{-}}$ for all $\theta$. It generally holds as $\mathbf{R}_y(2\theta) = e^{-i\theta \sigma_y}$ in which $\sigma_y$ has two eigenvalues of $\pm 1$. Thus, without loss of generality, we can represent the unitary $\mathbf{S}(\mathbf{x})$~\eqref{eqn: multimodal_unitary} with respect to the basis $\mathcal{B}_{+} = \{\ket{j}\ket{v^{+}}\ |\ j \in [0, MD] \}$:
\begin{equation}
\mathbf{S}(\mathbf{x}) = \begin{bmatrix}
        1 & & & \\
        & e^{i\phi_1} & & \\
        & & \ddots & \\
        & & & e^{i\phi_{MD}}\\
    \end{bmatrix},
\label{eqn: qubitization}
\end{equation}
and $\mathcal{H}_{+} = \text{span}(\mathcal{B}_{+})$ is an invariant subspace of $\mathbf{S}(\mathbf{x})$. On the other hand, the parameterized unitary $\mathbf{U}_p$ only performs on the index register such that for all $j\in [0, MD]$:
$$
\mathbf{U}_p\ket{j}\otimes\ket{v^{+}} = (\mathbf{U}_p\ket{j})\otimes \ket{v^{+}}.
$$
Then, the action of $\mathbf{U}_p$ only creates the superposition of the computational basis $\{\ket{j}\}$ without creating entanglement with the second register. Therefore, the $\mathbf{F}_{P}(\mathbf{x})$ is indeed performing the transformation on the eigenvalues of $\{e^{i0},e^{i\phi_1}, \dots, e^{i\phi_{MD}} \}$ of the shared eigenvector of $\ket{v^{+}}$ within the space $\mathcal{H}_{+}$. 

Before going to the full proof, we start with a toy example of two variables to see how our construction~\eqref{eqn: mqsvt} could generate polynomial functions. For convenience, we denote 
$$
\mathbf{S}_2(\mathbf{x}) := \begin{bmatrix}
        \mathbf{x}_1 &  \\
        & \mathbf{x}_2 \\
    \end{bmatrix},
$$ 
where $\mathbf{x}_1, \mathbf{x}_2 \in \mathbb{T}$, with $\mathbb{T} = \{{x} \in \mathbb{C}: |{x}| = 1\}$ is the complex unit circle. Thus, the unitary $\mathbf{U}_p$ can be seen as an arbitrary single-qubit operator. Without loss of generality, we can represent 
$$
\mathbf{U}_p = \begin{bmatrix}
    e^{i(\theta^{p}_1+\theta^{p}_2)}\cos(\theta^{p}_3) & e^{i\theta^{p}_2}\sin(\theta^{p}_3) \\
    e^{i\theta^{p}_1}\sin(\theta^{p}_3) & -\cos(\theta^{p}_3)
\end{bmatrix},
$$
which is the generic form to represent any single-qubit unitary \cite{nielsen2001quantum}. Then, we consider the quantum circuit $\mathbf{F}_2(\mathbf{x})$ with $P=2$ as:
\begin{equation}
    \mathbf{F}_2(\mathbf{x}) = \mathbf{U}_0.\mathbf{S}_2(\mathbf{x}).\mathbf{U}_1.\mathbf{S}_2(\mathbf{x}).\mathbf{U}_2 =\begin{bmatrix}
    A(\mathbf{x}) & B(\mathbf{x}) \\
   C(\mathbf{x}) & D(\mathbf{x}) \\
\end{bmatrix},
\end{equation}
with 
\begin{align}
A(\mathbf{x}) &:= w^1_1(\boldsymbol{\theta}) \mathbf{x}^2_1 + w^1_2(\boldsymbol{\theta}) \mathbf{x}_1\mathbf{x}_2 + w^1_3(\boldsymbol{\theta}) \mathbf{x}^2_2, \\
   B(\mathbf{x}) &:= w^2_1(\boldsymbol{\theta}) \mathbf{x}^2_1 + w^2_2(\boldsymbol{\theta}) \mathbf{x}_1\mathbf{x}_2 + w^2_3(\boldsymbol{\theta}) \mathbf{x}^2_2,\\
    C(\mathbf{x}) &:=  w^3_1(\boldsymbol{\theta}) \mathbf{x}^2_1 + w^3_2(\boldsymbol{\theta}) \mathbf{x}_1\mathbf{x}_2 + w^3_3(\boldsymbol{\theta}) \mathbf{x}^2_2,\\
D(\mathbf{x}) &:= w^4_1(\boldsymbol{\theta}) \mathbf{x}^2_1 + w^4_2(\boldsymbol{\theta}) \mathbf{x}_1\mathbf{x}_2 + w^4_3(\boldsymbol{\theta}) \mathbf{x}^2_2, 
\end{align}
where $\boldsymbol{\theta}$ is the set of all parameters of $\{(\theta^{p}_1, \theta^{p}_2, \theta^{p}_3)\}_{p=1}^{3}$ and $\{w^i_j\}$ are some functions of $\boldsymbol{\theta}$. We can see that the circuit $\mathbf{F}_2(\mathbf{x})$ generate the homogeneous 2-degree polynomials functions of two variables $\mathbf{x}_1, \mathbf{x}_2$. However, in the general case, with the element of $1$ in~\eqref{eqn: qubitization}, the quantum circuit can represent any monomials.

We now formalize the theoretical foundation of our construction in the following theorem.

\begin{figure*}
\begin{minipage}{\textwidth}
  \centering
  \begin{subfigure}[b]{0.3\textwidth}
        \includegraphics[width=\linewidth]{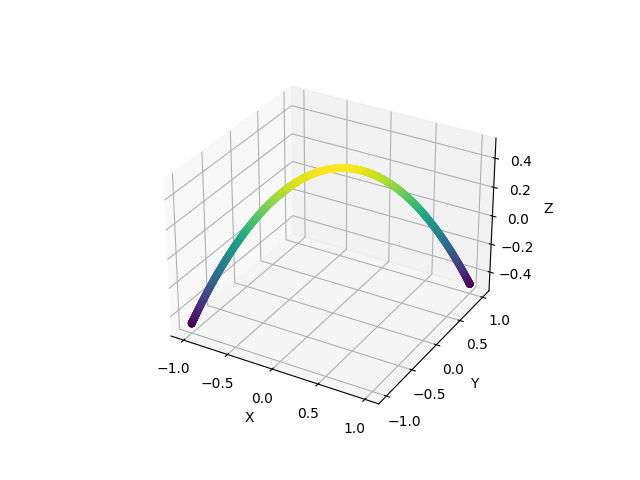}
        \caption{$P=1$}
        \label{fig:depth1}
    \end{subfigure}
    \hfill
    \begin{subfigure}[b]{0.3\textwidth}
        \includegraphics[width=\linewidth]{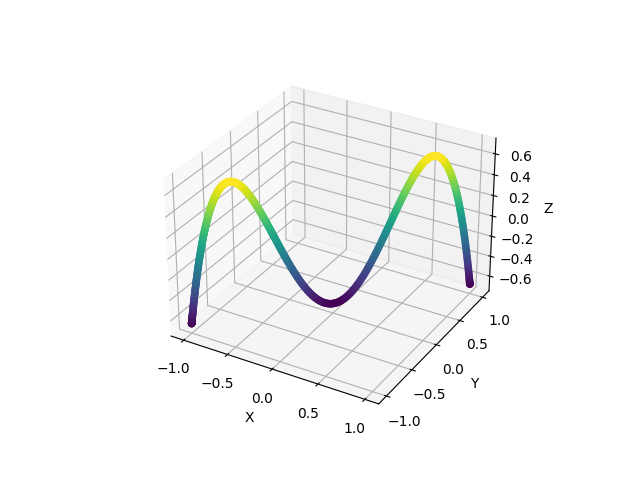}
        \caption{$P=2$}
        \label{fig:depth2}
    \end{subfigure}
    \hfill
    \begin{subfigure}[b]{0.3\textwidth}
        \includegraphics[width=\linewidth]{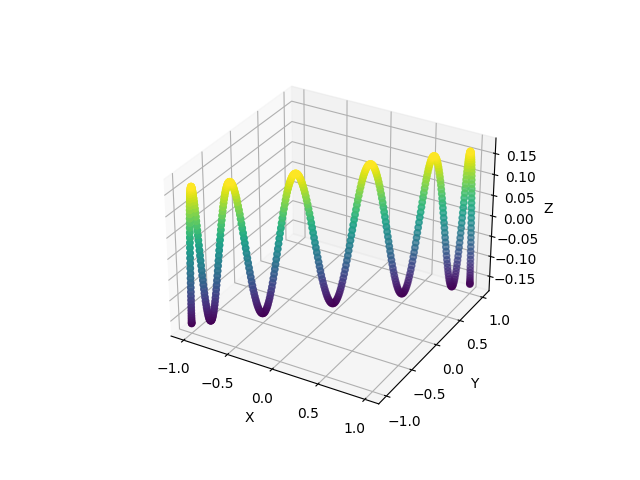}
        \caption{$P=6$}
        \label{fig:depth6}
    \end{subfigure}
    \caption{Outputs of Quantum Fusion Layer on $\mathbf{x}_1 = e^{-i\theta_1}, \mathbf{x}_2 = e^{-i\theta_2}$ for $\theta_1, \theta_2 \in [0, 2\pi]$ with increasing depth $P=1, 2, 6$.}
    \label{fig:qfl_output}
\end{minipage}
\end{figure*}

\begin{thrm}[Expressivity]
    For all \( P \in \mathbb{N} \), there exists a family of unitaries \( \{\mathbf{U}_p\}_{p=0}^{P} \subset SU(MD+1) \) such that the product structure defined in~\eqref{eqn: mqsvt} yields a matrix-valued function \( \mathbf{F}_{P}(\mathbf{x}) \) satisfying:
    \begin{enumerate}
        \item \( \mathbf{F}_{P}(\mathbf{x}) \) is a matrix-valued polynomial of total degree at most \( P \);
        \item \( \mathbf{F}_{P}(\mathbf{x}) \in SU(2(MD+1)) \) for all \( \mathbf{x} = (\mathbf{x}_1, ..., \mathbf{x}_{MD}) \in \mathbb{T}^{MD} \);
        \item \( \det(\mathbf{F}_{P}(\mathbf{x})) = 1 \).
    \end{enumerate}
    Here, \( \mathbb{T} = \{x \in \mathbb{C} : |x| = 1\} \) denotes the complex unit circle.
    \label{thrm: main_thrm}
\end{thrm}
The proof is provided in Appendix~\ref{app: proof_thrm1}. Theorem~\ref{thrm: main_thrm} establishes that the quantum circuit \( \mathbf{F}_P(\mathbf{x}) \), as defined in~\eqref{eqn: mqsvt}, can generate matrix-valued polynomial functions of arbitrary degree \( P \) over the torus \( \mathbb{T}^{MD} \), while preserving unitarity and determinant constraints. This expressive capability implies that, in principle, the circuit can approximate any continuous function on \( \mathbb{T}^{MD} \) arbitrarily well as \( P \to \infty \), consistent with the Stone–Weierstrass theorem \cite{de1959stone}. Figure~\ref{fig:qfl_output} visualizes the output of the quantum fusion layer on two variables $\mathbf{x}_1 = e^{-i\theta_1}, \mathbf{x}_2 = e^{-i\theta_2}$ for $\theta_1, \theta_2 \in [0, 2\pi]$ with different depths $P = 1,2,6$. The outputs correspond to degree-$P$ complex polynomials on $\mathbb{T}^2$.

We now consider the \textit{parameter efficiency} of our model: how many tunable parameters are needed to realize its expressive power? Let \( K \) be the number of parameters per unitary block \( \mathbf{U}_p \), and let \( P \) denote the circuit depth. The total number of parameters thus scales with \( K \cdot P \).

Our goal is for \( \mathbf{F}_P(\mathbf{x}) \) to represent any degree-\( P \) multivariate polynomial over the domain \( \mathbb{T}^{MD} \). Geometrically, this requires each \( \mathbf{U}_p \) to cover a sufficiently large subspace of the Hilbert space ideally the entire Bloch sphere to enable uniform approximation. While full coverage requires \( \mathcal{O}(2^n) \) degrees of freedom for \( n \)-qubit systems, practical quantum models often only need \textit{approximate unitary 2-designs}, which replicate low-order Haar moments. Cleve et al.~\cite{cleve2015near} showed that such 2-designs can be implemented with \( \Tilde{\mathcal{O}}(n) \) gates and logarithmic depth. In our setting, this yields a parameter bound of \( K = \mathcal{O}(Md) \), ensuring efficient yet expressive quantum circuits.
\begin{corollary}[Parameter Scaling]
Let \( \mathbf{F}_P(\mathbf{x}) \) be the quantum circuit defined in~\eqref{eqn: mqsvt}. For any \( P > 0 \), the circuit can approximate any matrix-valued polynomial of degree at most \( P \) on \( \mathbb{T}^{Md} \), under the constraint \( \det(\mathbf{F}_P(\mathbf{x})) = 1 \). The required parameter complexity depends on the expressivity of the unitary blocks \( \mathbf{U}_p \) as follows:

\begin{itemize}
    \item[(i)] If each \( \mathbf{U}_p \) can densely cover the Bloch sphere of an \( n \)-qubit subsystem, then \( K = \mathcal{O}(MD) \), giving total complexity \( \Tilde{\mathcal{O}}(P \cdot MD) \).
    
    \item[(ii)] If each \( \mathbf{U}_p \) is drawn from an approximate unitary 2-design, the same approximation can be achieved with \( \Tilde{\mathcal{O}}(P \cdot \mathrm{poly}(\log(MD))) \) parameters.
\end{itemize}

\label{cor:expressivity_tradeoff}
\end{corollary}
This result highlights a fundamental expressivity–efficiency tradeoff in QFL: highly expressive unitary blocks enable dense polynomial approximation at linear parameter cost, while approximate 2-designs reduce parameter growth even further by leveraging randomness. In either case, the resulting scaling is polynomial, contrasting sharply with the exponential complexity of classical full-rank tensor networks.  

Next, we present the computational complexity of QFL, analyzing how these parameter-scaling guarantees translate into overall gate complexity when state preparation, parameterized circuits, and randomized measurements are combined. It is given in the following theorem:
\begin{thrm}
    [Computational Complexity]\label{thrm: computational_complexity}
    Fix $M$ modalities, each with feature dimension $D$. Consider a QFL architecture consisting of a state-preparation circuit for $\mathbf{S}(\mathbf{x})$, followed by a parameterized quantum circuit of depth $P$ with blocks $\{U_p\}_{p=1}^P$, where the gate counts of the blocks are denoted $\{|U_p|\}_{p=1}^P$. Suppose the output layer involves randomized measurement of $H$ observables, each estimated to additive error at most $\epsilon$ with success probability at least $1-\delta$. Then the total gate complexity required to estimate the resulting $H$-dimensional expectation vector is
\[
\tilde{\mathcal{O}}\!\left((MD \;+\; P \cdot \max_{p}|U_p|)\cdot H \cdot \frac{1}{\epsilon^2}\right),
\]
where $\tilde{\mathcal{O}}(\cdot)$ hides logarithmic factors in $1/\delta$. 
\end{thrm}
We present the proof in Appendix~\ref{app: proof_thrm2}.

\subsection{Advantages over Low-rank Tensor Methods}

As a first step toward a formal quantum advantage in multimodal fusion, we ground our analysis in tensor-based techniques. which are the natural baseline for head-to-head comparison between quantum and classical models (cf.~\cite{Rudolph_2023}). However, a universal theoretical framework for principled comparison between classical and quantum learning models, particularly in the multimodal setting, remains elusive. Accordingly, in this section, we establish a query-complexity separation between QFL and low-rank tensor approximation methods (e.g., LMF), and leave broader comparisons as an open problem. 

From Corollary~\ref{cor:expressivity_tradeoff}, we observe that both low-rank tensor approximations and parameterized quantum circuits achieve \textit{polynomial parameter scaling}. This naturally raises a key question:
\begin{center}
    \textit{Where does the quantum advantage arise?}
\end{center}
To address this, we construct a function for which classical low-rank methods based on separable representations fail to match the \textit{query complexity} of our quantum approach. Specifically, we focus on the CP format, which approximates multivariate functions as sums of products of univariate functions~\eqref{eqn: cp_format}.  

\begin{prob}
Consider \( \mathbf{x}_1 = e^{i\theta_1} \), \( \mathbf{x}_2 = e^{i\theta_2} \). Approximate the function:
\[
    f(\mathbf{x}_1, \mathbf{x}_2) =
    \begin{cases}
        0 & \text{if } \{\theta_1, \theta_2\} \in \{\{0, \pm \pi/2\}, \{\pm \pi/2, 0\}\}, \\
        1 & \text{if } 4\cos^2(\theta_1)\cos^2(\theta_2) = 1.
    \end{cases}
\]
\label{prob: 1}
\end{prob}

\begin{thrm}
    Consider \( \mathbf{x}_1 = e^{i\theta_1} \), \( \mathbf{x}_2 = e^{i\theta_2} \), there exists a QFL construction can solve the Problem~\ref{prob: 1} exactly using six joint queries to \( \mathbf{S}(\mathbf{x}_1, \mathbf{x}_2) \)  
    \label{thrm: adv}
\end{thrm}
The problem, inspired by quantum multi-channel discrimination~\cite{rossi2022quantum}, can be solved with only \emph{six joint queries} to the oracle \( \mathbf{S}(\mathbf{x}_1, \mathbf{x}_2) \) as Theorem~\ref{thrm: adv}. The proof are provided in Appendix~\ref{app: prob_1}. In contrast, classical low-rank tensor network-based models, which must query $\mathbf{x}_1$ and $\mathbf{x}_2$ independently as in~\eqref{eqn: cp_format}, cannot achieve the same performance within comparable query complexity~\cite{rossi2022quantum}. While a full-rank tensor representation could in principle close this gap, it would incur exponential parameter growth, thereby undermining the motivation for low-rank approximations. This establishes a formal separation between QFL and low-rank tensor fusion, and importantly, the advantage persists even when both parties accessing $\mathbf{x}_1$ and $\mathbf{x}_2$ are quantum-powered but lack entanglement.  

Although this separation highlights a promising theoretical guarantee for quantum multimodal fusion, its direct practical relevance is limited. As in other rigorously established quantum advantages for tasks such as concept learning and kernel methods~\cite{liu2021rigorous, lewis2025quantum, bshouty1995learning}, the result is primarily worst-case and may not manifest empirically on standard benchmark datasets. Furthermore, current quantum hardware constraints, including limited qubit counts, noise, and shallow circuit depth, make it difficult to scale quantum learning models to large or complex regimes~\cite{cerezo2022challenges}. Consequently, identifying real-world multimodal settings where quantum learning can surpass classical approaches remains an open and active research direction.  

Given these considerations, our experiments should be interpreted as a \emph{proof of concept} for QFL rather than a demonstration of practical quantum advantage. In the next section, we empirically compare QFL with strong classical baselines across real-world multimodal datasets, with the goal of assessing its feasibility and behavior under realistic data conditions.  

\section{Experiments}\label{sec: exp}
\begin{table*}[t]
\centering
\caption{Performance comparison across datasets with varying modality counts. Arrows indicate the desired direction for each metric: ↑ means \textbf{higher is better}, ↓ means \textbf{lower is better}. Best results per dataset are shown in \textbf{bold}. Metrics that are not applicable or not reported for certain methods are denoted with ``--''.}
\label{tab:results}
\resizebox{\textwidth}{!}{%
\begin{tabular}{l||l||c||c||c||c}
\toprule
\textbf{Dataset} & \textbf{Method} & \textbf{Trainable Params} $\downarrow$ & \textbf{ Accuracy (\%)} $\uparrow$ & \textbf{F1-score} $\uparrow$& \textbf{ROC AUC} $\uparrow$\\
\midrule
\midrule
\multirow{5}{*}{{Multimodal Entailment (2 Modalities)}} 
& Concatenate                    & 1,248,711 & 84.29 & 0.7709 & -- \\
& MFB \cite{yu2017multi}           & 2,564,611                & 77.14 & 0.7945 & --\\
& \textbf{QFL (P=1) }              & \textbf{1,249,434}       & \textbf{86.42} & \textbf{0.8376 } & --\\
& QFL (P=3)   & 1,249,736 & 86.42 & 0.8156 & --\\
& QFL (P=5)   & 1,250,038 & 79.28 & 0.7983 & --\\
\midrule
\midrule
\multirow{3}{*}{{PTB-XL (3 Modalities)}}
& LMF \cite{liu2018efficient}                       & 1,780,823    & -- & -- & 0.851\\
& \textbf{QFL (P=5)}                    & \textbf{183,321}  & -- & -- & \textbf{0.859}\\
\midrule
\midrule
\multirow{3}{*}{{PTB-XL (5 Modalities)}}
& LMF   \cite{liu2018efficient}                      & 2,867,287 & -- & -- & 0.846\\
& \textbf{QFL (P=5)}                      & \textbf{204,980} & -- & -- & \textbf{0.850} \\
\midrule
\midrule
\multirow{3}{*}{{PTB-XL (12 Modalities)}}
& LMF     \cite{liu2018efficient}               & 6,669,911        & -- & -- & 0.500\\
& \textbf{QFL (P=5)}                    & \textbf{280,399}  & -- & -- & \textbf{0.887} \\
\midrule
\midrule
\multirow{4}{*}{{Traffic-LA (207 Modalities)}}
& GCN (1 layer) \cite{kipf2017semisupervisedclassificationgraphconvolutional}                      & \textbf{241,309} & 88.46 & \textbf{0.8375} & 0.8189 \\
& GCN (2 layers)  \cite{kipf2017semisupervisedclassificationgraphconvolutional}                    &  241,381 & 88.29 & 0.8328 & 0.8130\\
& QFL (P=1)                     & 241,392 & 87.76 & 0.8258 & 0.9078\\
& \textbf{QFL (P=2)}           & 241,595 & \textbf{88.46} & 0.8321 & \textbf{0.9151} \\
& QFL (P=3)           & 241,798& 87.76 & 0.8267 & 0.8708 \\
\bottomrule
\end{tabular}
}
\end{table*}


In this section, we evaluate QFL's performance using TensorFlow-Quantum~\cite{broughton2021tensorflowquantumsoftwareframework}, beginning with a detailed description of the experimental setup, followed by comprehensive results and analysis across benchmarks. The evaluation is carried out on a computer with an NVIDIA RTX A5500 GPU. More experimental details are presented in Appendix~\ref{app: exp_design}.

\subsection{Experimental Methodology} \label{sec: dataset&baseline}

\paragraph{Datasets:} We evaluate our method on three multimodal datasets spanning a range of domains and modality configurations. (i) \textbf{Multimodal Entailment} \cite{ilharco-etal-2021-recognizing}, combining vision and text (V+T), (ii) \textbf{PTB-XL} \cite{wagner2020ptb}, with 12-lead ECG signals treated as separate modalities, and (iii) \textbf{Traffic-LA} \cite{li2018diffusionconvolutionalrecurrentneural}, containing 207 traffic sensor streams. These benchmarks cover a wide spectrum of modality counts and interaction complexity.

\paragraph{Baselines:} To assess QFL, we compare it to strong classical baselines. These include low-order fusion methods such as \textbf{Concatenation} and \textbf{MFB }\cite{yu2017multi}, as well as scalable approaches like \textbf{LMF} \cite{liu2018efficient} and \textbf{GCNs} \cite{kipf2017semisupervisedclassificationgraphconvolutional}. To ensure fair comparison, we align pre- and post-fusion dimensionalities across QFL and baselines (except Concatenation). These configurations are detailed in Table~\ref{tab:fusion_dim}.

\begin{table}[t]
\centering
\caption{Dimensionality settings of the pre- and post-fusion layers across datasets. The pre-fusion layer dimensionality reflects the feature dimension assigned to each modality before fusion.}
\label{tab:fusion_dim}
\begin{tabular}{l||c||c}
\toprule
\textbf{Dataset} & \textbf{Fusion IN} & \textbf{Fusion OUT} \\
\midrule
Multimodal Entailment & 256 & 512 \\
PTB-XL & 64 & 256 \\
Traffic-LA & 16 & 16 \\
\bottomrule
\end{tabular}
\end{table}

\paragraph{Evaluation Metrics: } We evaluate the performance of all models using standard classification metrics, including \textbf{accuracy}, \textbf{F1-score}, and \textbf{ROC AUC}, tailored to the nature of each dataset. For class-imbalanced tasks such as Multimodal Entailment, we report the F1-score as a weighted average across classes to ensure fair evaluation. In contrast, for PTB-XL and Traffic-LA, we use the ROC AUC as the primary metric, which provides a more informative measure of discriminative ability across varying classification thresholds. We also report the number of \textbf{trainable parameters} to evaluate scalability, with lower values preferred. Higher metric values indicate better performance.

\paragraph{Experimental Details:} For \textbf{Multimodal Entailment} (2 modalities), we evaluate QFL with $P{=}1,3,5$ to examine the impact of fusion depth. For \textbf{PTB-XL}, we evaluate the scalability of QFL and {LMF} on 3, 5, and 12-lead subsets, using QFL with fixed $P{=}5$ to capture high-order dependencies and assess performance as modality count grows. Since neither of these datasets exhibits an inherent graph structure, we do not include {GCNs} in their comparisons. For \textbf{Traffic-LA}, which includes 207 modalities, we exclude LMF due to trainability issues and compare QFL with {GCNs}. To accommodate limited simulation resources, we evaluate shallow configurations with $P{=}1,2,3$ against 1- and 2-layer GCNs.

\subsection{Results and Discussion} \label{sec: results}
Table~\ref{tab:results} summarizes the performance of the Quantum Fusion Layer (QFL) across three multimodal benchmarks in comparison to the baselines.

\paragraph{Comparison with Baselines: } On \textbf{Multimodal Entailment}, QFL with polynomial degree $P{=}1$ achieves the highest accuracy (86.42\%) and matches the best F1-score (0.8376), outperforming both Concatenation and MFB despite using fewer trainable parameters. Increasing the polynomial degree to $P{=}3$ slightly improves the F1-score to 0.8156, while $P{=}5$ shows mild overfitting, reflected in a drop in both accuracy and F1. QFL's advantages become more pronounced in high-modality regimes. On \textbf{PTB-XL}, QFL maintains stable AUCs as the number of modalities increases, achieving 0.859 with 3 leads, 0.850 with 5 leads, and a strong 0.887 with all 12 leads. In contrast, LMF collapses to an AUC of 0.500 when the number of modalities goes beyond 5 due to vanishing gradient. Notably, QFL achieves these results with significantly fewer parameters than LMF. In the best-case scenario (with 12 input leads), QFL uses only $5\%$ of the parameters required by LMF while achieving a $76\%$ improvement in accuracy.
On the \textbf{Traffic-LA} dataset with 207 modalities, QFL consistently outperforms GCN baselines. 
Notably, QFL with $P{=}2$ achieves the highest AUC (0.9151), representing a $12\%$ improvement over GCN, along with the best F1-score (0.8298). It also maintains competitive accuracy ($88.67\%$) while using a comparable number of parameters to GCN. The performance remains stable across different $P$. The results highlight QFL’s scalability and effectiveness in modeling cross-modal dependencies, exhibiting strong generalization, especially in scenarios with a large number of modalities.



\paragraph{Impact of Quantum Modeling:} QFL avoids the limitations of tensor decomposition methods like LMF, which approximate high-order interactions via fixed-rank structures. While effective for small \( M \), LMF suffer from vanishing gradients due to the multiplicative interaction across modalities. This can only be mitigated if the rank $R$ grows exponentially with $M$, making the approach computationally infeasible at scale. In contrast, QFL uses parameterized quantum circuits to implicitly encode high-order dependencies without exponential parameter growth. Moreover, unlike GCN-based models, which rely on local message passing and require deep architectures to capture long-range or global modality interactions, QFL captures global dependencies even with shallow circuits. This allows QFL to outperform GCNs in shallow settings.

\begin{figure}
    \centering
    \includegraphics[width=0.4\linewidth]{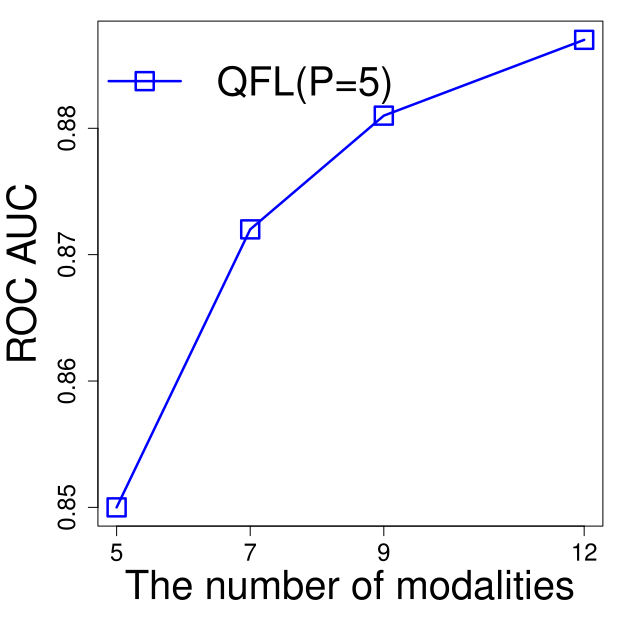}
    \caption{\textbf{Impacts of the modality on QFL.} QFL performance improves as the number of input modalities increases.}
    \label{fig: qfl_increase_num}
\end{figure}

\paragraph{Ablation Study on Modality Scaling: } We assess QFL’s robustness to increasing modality count on the PTB-XL dataset using 5, 7, 9, and 12 ECG leads. As shown in Figure~\ref{fig: qfl_increase_num}, QFL’s ROC AUC improves consistently, from 0.850 (5 leads) to 0.887 (12 leads), indicating stable and enhanced performance with more modalities. This suggests QFL effectively leverages additional modality-specific information without the significant gradient degradation or overfitting commonly seen in classical methods~\cite{wang2020makes}.

\section{Conclusion and Open Problems}\label{sec: open_prob}

In this work, we introduced the \emph{Quantum Fusion Layer} (QFL), a hybrid quantum--classical framework for multimodal learning. QFL leverages parameterized quantum circuits to model entangled, non-separable feature interactions across modalities without incurring the exponential parameter growth faced by classical full-tensor fusion. Our theoretical analysis established two key results: (i) an expressivity theorem showing that QFL can represent multivariate polynomials with only linear parameter growth, and (ii) a query-complexity separation demonstrating that QFL achieves representational power unattainable by low-rank tensor fusion methods under comparable resource constraints.  

Beyond theory, we benchmarked QFL against classical low-rank and graph-based fusion baselines on diverse multimodal tasks. While no separation result is currently known against these more flexible architectures, QFL consistently outperformed classical baselines in simulation, with its strongest improvements appearing in high-modality regimes where its theoretical strengths are most relevant. These results highlight the potential of QFL as both a principled and scalable fusion mechanism.  

At the same time, our findings should be viewed as a proof of concept rather than a demonstration of practical quantum advantage. The limitations of present-day quantum hardware, such as restricted qubit counts, shallow depth, and noise, pose significant challenges to scaling QFL to large and realistic settings. Designed with fault-tolerant architectures in mind, QFL opens a path toward future quantum fusion models, while its effectiveness on near-term intermediate-scale devices remains an important open question.  

Overall, this work contributes the first rigorous separation theorem in multimodal fusion together with empirical evidence of superior performance in high-modality tasks. By bridging provable expressivity with practical feasibility, QFL provides a foundation for further exploration of quantum techniques in multimodal learning and motivates the search for real-world datasets and hardware regimes where quantum fusion can demonstrate tangible advantage.

\paragraph{Acknowledgment.} This work was initiated while Tuyen N. was at the University of Aizu. The author thanks M\'aria Kieferov\'a and Gabriel Waite for helpful discussions related to this paper. Tuyen N. is supported by a scholarship from the  Sydney Quantum Academy, PHDR06031.
\medskip 

\printbibliography
\clearpage

\begin{appendix}

\section{Theoretical Proofs}
\subsection{Proof of Theorem \ref{thrm: main_thrm}} \label{app: proof_thrm1}
For convenience, we restate Theorem~\ref{thrm: main_thrm} here.
\begin{thrm}
    For all \( P \in \mathbb{N} \), there exists a family of unitaries \( \{\mathbf{U}_p\}_{p=0}^{P} \subset SU(MD+1) \) such that the product structure defined in (11) yields a matrix-valued function \( \mathbf{F}_{P}(\mathbf{x}) \) satisfying:
    \begin{enumerate}
        \item \( \mathbf{F}_{P}(\mathbf{x}) \) is a matrix-valued polynomial of total degree at most \( P \);
        \item \( \mathbf{F}_{P}(\mathbf{x}) \in SU(2(MD+1)) \) for all \( \mathbf{x} = (\mathbf{x}_1, ..., \mathbf{x}_{MD}) \in \mathbb{T}^{MD} \);
        \item \( \det(\mathbf{F}_{P}(\mathbf{x})) = 1 \).
    \end{enumerate}
    Here, \( \mathbb{T} = \{x \in \mathbb{C} : |x| = 1\} \) denotes the complex unit circle.
\end{thrm}
\proof{
    The proof proceeds by induction on the polynomial degree \( P \).

    For \( P = 0 \), the function \( \mathbf{F}_0(\mathbf{x}) = \mathbf{U}_0 \) is constant, as it does not depend on \( \mathbf{x} \), and clearly belongs to \( SU(2(MD+1)) \) with determinant one. All three properties are trivially satisfied.

    Suppose the theorem holds for degree \( P = t-1 \), with the associated function \( {\mathbf{F}_{t-1}}(\mathbf{x}) \). We define:
    \[
        \mathbf{F}_t(\mathbf{x}) = {\mathbf{F}_{t-1}}(\mathbf{x})\cdot \mathbf{S}(\mathbf{x}) \cdot \mathbf{U}_t ,
    \]
    where \( \mathbf{U}_t \in SU(MD+1) \) is independent of \( \mathbf{x} \), and \( \mathbf{S}(\mathbf{x}) \in SU(2(MD+1)) \) is a shift gate or diagonal operator depending linearly on \( \mathbf{x} \). Since \( {\mathbf{F}_{t-1}}(\mathbf{x}) \) is a degree$-(t-1)$ polynomial and \( \mathbf{S}(\mathbf{x}) \) has degree one, their product is a polynomial of degree at most \( t \). Furthermore, the product of unitaries is unitary, and determinants multiply, so all three properties are preserved.

    Finally, the embedding of each \( \mathbf{U}_p \) as \( \mathbf{U}_p \otimes \mathbb{I} \) within the larger space \( SU(2(MD+1)) \), together with \( S \in SU(2(MD+1)) \), ensures that the output remains in \( SU(2(MD+1)) \) for all \( \mathbf{x} \in \mathbb{T}^{MD} \).}

\subsection{Proof of Theorem~\ref{thrm: computational_complexity}}\label{app: proof_thrm2}
The cost analysis of QFL can be decomposed into three modules: \emph{state preparation}, \emph{parameterized quantum circuit}, and \emph{measurement}. We now analyze the cost of each module step by step.  

It is straightforward to establish the computational complexity of state preparation and the parameterized circuit. For state preparation, as shown in Section~\ref{sec: approach}, preparing $\mathbf{S}(\mathbf{x})$ requires a quantum circuit of $\mathcal{O}(MD)$ gates acting on $\lfloor \log(MD+1) \rfloor + 1$ qubits. For the parameterized quantum circuit, since our design deliberately leaves the choice of ansatz general, the cost is given by $\mathcal{O}(P \cdot \max_{p} |U_p|)$, where $|U_p|$ denotes the number of gates implementing the unitary $U_p$.  

We now turn to the cost of the measurement layer. We consider $H$ observables $\{ \mathcal{M}_i \}_{i=1}^{H}$ corresponding to randomized measurements in the planes spanned by Pauli pairs, as described in Section~\ref{sec: approach}. Since each $\mathcal{M}_i$ is a linear combination of two Pauli operators, the following lemma applies.  

\begin{lmma}
Consider $\mathcal{M} = \sum_{i=1}^{m} c_i P_i$ with $P_i \in \{\pm I, \pm \sigma_x, \pm \sigma_y, \pm \sigma_z\}$ and coefficients $|c_i|\leq 1$. Suppose a circuit run consists of: preparing $\ket{\psi}$, rotating into the eigenbasis of a chosen Pauli $P_i$, measuring in the computational basis, and recording $X_i \in \{\pm 1\}$ with $\mathbb{E}[X_i] = \bra{\psi}P_i\ket{\psi}$. Then, to obtain an estimator
\[
\hat{\mu} = \sum_{i=1}^m c_i\,\overline{X}_i, 
\quad 
\overline{X}_i = \frac{1}{N_i}\sum_{t=1}^{N_i} X_i^{(t)},
\]
satisfying
\[
|\hat{\mu} - \bra{\psi}\mathcal{M}\ket{\psi}| \leq \epsilon
\]
with probability at least $1-\delta$, it suffices to run the circuit
\[
N = \mathcal{O}\!\left( \frac{m}{\epsilon^2}\log\frac{1}{\delta} \right)
\]
times.
\end{lmma}

\begin{proof}
Unbiasedness follows from linearity of expectation:
\[
\mathbb{E}[\hat{\mu}] = \sum_{i=1}^m c_i \,\mathbb{E}[\overline{X}_i] 
= \sum_{i=1}^m c_i \,\langle \psi | P_i | \psi \rangle 
= \langle \psi | \mathcal{M} | \psi \rangle.
\]
Since $X_i \in \{\pm 1\}$, we have $\operatorname{Var}(X_i)\leq 1$, hence $\operatorname{Var}(\overline{X}_i)\leq 1/N_i$. Independence across $i$ gives
\[
\operatorname{Var}(\hat{\mu}) 
= \sum_{i=1}^m c_i^2 \,\operatorname{Var}(\overline{X}_i) 
\leq \sum_{i=1}^m \frac{c_i^2}{N_i}.
\]
Optimizing $\{N_i\}$ subject to $\sum_i N_i = N$ yields $N_i^\star = N\,|c_i|/\Lambda_1$ with $\Lambda_1 = \sum_{i=1}^m |c_i| \leq 1$, giving $\operatorname{Var}(\hat{\mu}) \leq m/N$. Finally, Hoeffding’s inequality implies that $N=\mathcal{O}(m\epsilon^{-2}\log(1/\delta))$ samples suffice.
\end{proof}

Thus, estimating each observable requires $\tilde{\mathcal{O}}(m\epsilon^{-2})$ circuit runs, and over $H$ observables, the measurement cost is $\tilde{\mathcal{O}}(H\epsilon^{-2})$ runs.  

\paragraph{Putting it all together.} Combining the costs of state preparation, parameterized circuits, and measurement, the overall gate complexity of QFL is
\[
\tilde{\mathcal{O}}\!\left((MD + P \cdot \max_{p}|U_p|)\cdot H \cdot \frac{1}{\epsilon^2}\right).
\]

\subsection{Proof of Theorem~\ref{thrm: adv}} \label{app: prob_1}
The function in Problem~\ref{prob: 1} is inspired by Problem~3.1 in~\cite{rossi2022multivariable}, and it demonstrates a separation in query complexity between different access models: \textit{coherent} and \textit{incoherent} access models. Several recent frameworks explore quantum advantage in this setting \cite{aharonov2022quantum, huang2021information, rossi2022quantum}. We refer interested readers to these papers for a more detailed discussion of the separation.

Rossi \textit{et al.} \cite{rossi2022multivariable} establish the separation results for Problem~\ref{prob: 1} based on the fact that there exists a quantum algorithm with QSP structure accessing $3$ joint oracles of $\mathbf{x}_1$ and $\mathbf{x}_2$ or 6 queries to $\mathbf{x}_1$, $\mathbf{x}_2$ that could deterministically solve the problem with zero error (Lemma IV.4. \cite{rossi2021quantum}). On the other hand, no two parties, each grants access to $\mathbf{x}_1$ and $\mathbf{x}_2$ independently, connected only by classical communication and sharing no entanglement, could decide this problem with the same performance (Theorem II.1 \cite{rossi2022quantum}). Notably, this also holds even if the two parties are quantum-powered. 

Intuitively, this advantage arises because the structure of the discrimination problem cannot be decomposed into regions that depend only on individual inputs ($\theta_1$ and $\theta_2$). Instead, the function fundamentally relies on a joint property of both inputs, making it hard to resolve by local projections followed by classical combinations.

This key insight directly allows us to compare the quantum structure with low-rank tensor networks. The methods, like the CP decomposition, approximate the function as
\[
f(\mathbf{x}_1, \mathbf{x}_2) \approx \sum^R_\alpha g^{(1)}_\alpha(\mathbf{x}_1) \cdot g^{(2)}_\alpha(\mathbf{x}_2),
\]
which assumes independent access to each variable. Specifically, the inputs \( \mathbf{x}_1 \) and \( \mathbf{x}_2 \) are processed separately through distinct components (different functions $g_\alpha$) of the model. The expressive capacity of such models is governed by the rank \( R \). However, when \( R \) is low, these methods are inherently limited in their ability to capture non-separable joint behaviors, such as entanglement or interference. Consequently, they are unable to reproduce the correlations encoded in the function from Problem~\ref{prob: 1} using the same number of queries as models with coherent access.

Lastly, we establish a reduction from our construction of the state preparation operator \( \mathbf{S}(\mathbf{x}_1, \mathbf{x}_2) \) to the joint oracle formulation proposed in~\cite{rossi2022multivariable}. Their quantum signal processing (QSP)-based circuit for computing the function in Problem~\ref{prob: 1} takes the form:
\[
    \mathbf{F}(\mathbf{x}_1, \mathbf{x}_2) = e^{i \phi_0 \sigma_z} \prod_{k=0}^{2} e^{i\theta_1 \sigma_x} \cdot e^{i \phi_{2k+1}\sigma_z} \cdot e^{i \theta_2 \sigma_x} \cdot e^{i \phi_{2k+2}\sigma_z} ,
\]
where \( \phi_j = (-1)^j \pi/4 \) are fixed phase parameters. Note that \( \mathbf{F}(\mathbf{x}_1, \mathbf{x}_2) \) is a single-qubit operator acting jointly on the input parameters \( \theta_1 \) and \( \theta_2 \). The joint oracle used in this construction is:
\[
    e^{i\theta_1 \sigma_x} \cdot e^{i \phi_j\sigma_z} \cdot e^{i \theta_2 \sigma_x}, \quad \text{for } j \in \{1, \ldots, 6\}.
\]

We aim to reproduce this oracle within our Quantum Fusion Layer framework by appropriately selecting the unitary transformation \( \mathbf{U} \) and leveraging our block-encoded state preparation operator:
\[
    \mathbf{S}(\mathbf{x}_1, \mathbf{x}_2) =
    \begin{bmatrix}
        e^{i \theta_1 \sigma_y } & 0 \\
        0 & e^{i \theta_2 \sigma_y }
    \end{bmatrix},
\]
which acts on two qubits: a data register and an index register. To this end, consider the following operator sequence:
\begin{equation}
\begin{aligned}
&\Big( e^{i \frac{\pi}{4} \sigma_z} \, \mathbf{S}(\mathbf{x}_1, \mathbf{x}_2) \, e^{-i \frac{\pi}{4} \sigma_z} \Big) 
\cdot e^{i \phi_j \sigma_z} \cdot H  \\
& \quad \quad \quad \quad \quad \quad \quad \quad \quad \quad \cdot \Big( e^{i \frac{\pi}{4} \sigma_z} \, \mathbf{S}(\mathbf{x}_1, \mathbf{x}_2) \, e^{-i \frac{\pi}{4} \sigma_z} \Big) \\
&= \frac{1}{\sqrt{2}}
\begin{bmatrix}
    e^{i \theta_1 \sigma_x} \, e^{i \phi_j \sigma_z} \, e^{i \theta_1 \sigma_x} & e^{i \theta_1 \sigma_x} \, e^{i \phi_j \sigma_z} \, e^{i \theta_2 \sigma_x} \\
    e^{i \theta_1 \sigma_x} \, e^{i \phi_j \sigma_z} \, e^{i \theta_1 \sigma_x} & -e^{i \theta_2 \sigma_x} \, e^{i \phi_j \sigma_z} \, e^{i \theta_2 \sigma_x}
\end{bmatrix}.
\end{aligned}
\label{eqn:transform}
\end{equation}

Without loss of generality, we define \( \mathbf{U} = e^{-i \frac{\pi}{4} \sigma_z} \cdot e^{i \phi_j \sigma_z} \cdot H \cdot e^{i \frac{\pi}{4} \sigma_z} \) and absorb the global phase rotations \( e^{\pm i \frac{\pi}{4} \sigma_z} \) into adjacent layers. This construction yields a block-encoding of the desired joint oracle, showing that our architecture can realize the QSP-based multi-variable oracle in~\cite{rossi2022multivariable} with appropriate measurement, thereby enabling efficient implementation of Problem~\ref{prob: 1} within our Quantum Fusion Layer.

\section{ Experimental Design Details}\label{app: exp_design}
\subsection{ Datasets}

We evaluate QFL across three multimodal benchmarks spanning low to high modality regimes:

\begin{itemize}
    \item \textbf{Multimodal Entailment} \cite{ilharco-etal-2021-recognizing}: A classification task combining two modalities of \textit{text} and \textit{vision}. Similar to the task of \textit{text entailment}, the data set also tries to answer whether a given piece of information \textit{contradict} or \textit{imply} the other, but in a multi-modal way. Given pairs of (text-1, image-1) and (text-2, image-2), models are required to classify which label they belong to in \{entailment, no-entailment, contradiction\}. The dataset consists of $1000$ text–image pairs with a highly imbalanced distribution, over 80\% of the samples are labeled no-entailment. The data is split into $855$ training, $45$ validation, and $100$ test examples using stratified sampling to preserve class proportions. 
    
    
    \item \textbf{PTB-XL} \cite{wagner2020ptb}: A clinical ECG dataset with $12$ leads, each treated as an independent modality. The dataset contains $21837$ $10$-second $12$-lead recordings, annotated with three non-mutually exclusive categories of statements: \textit{diagnostic}, \textit{form}, and \textit{rhythm}. In this work, we combine all statements and form the classification task on a total of $71$ classes. We follow the data preprocessing in~\cite{strodthoff2020deep}.
    
    \item \textbf{Traffic-LA} \cite{li2018diffusionconvolutionalrecurrentneural}: The dataset captures traffic speed from 207 loop detectors across Los Angeles at 5‑minute intervals over a period of approximately three months, yielding over 6.5 million data points in total. Here, we follow the data preprocessing step in~\cite{ma2023federated}, where each sample consists of a $12$-timestep sequence from $207$ traffic sensors. This yields a total of $2856$ samples, each with a single feature (traffic speed), two output classes (e.g., congestion vs. no congestion), and a predefined spatial graph connecting the sensors. No missing values are present after preprocessing.
\end{itemize}
All datasets follow train/validation/test splits as the baselines.

\subsection{ Model Architecture}

\paragraph{Unimodal Encoder.} For all experiments, we employ the same unimodal encoders across both classical baselines and QFL. The details of unimodal corresponding to each task are described as follows:
\begin{itemize}
    \item \textbf{Multimodal Entailment:} We use a pretrained BERT-base model \cite{devlin2019bertpretrainingdeepbidirectional} to encode text, extracting the token embedding and projecting it to $256$ dimensions. Images are processed using a pretrained ResNet-50 backbone, with the average pool output similarly projected to $256$ dimensions.
    \item \textbf{PTB-XL:}  Each ECG lead is processed independently using a different 1D CNN encoder. The encoder consists of two convolutional layers (with $32$ and $64$ filters, respectively), each followed by batch normalization. A final global average pooling layer outputs a $64$-dimensional embedding per lead.
    \item \textbf{Traffic-LA:} Each sensor's $12$-timestep historical input is encoded using a different LSTM-based model. The encoder first applies an LSTM layer to capture temporal dependencies, producing a hidden representation at each timestep. The outputs are then passed through a dense layer to aggregate over time, yielding a $16$-dimensional embedding per sensor.
\end{itemize}

\paragraph{Fusion Techniques. } The specification of each fusion technique used in our experiment is provided as follows:
\begin{itemize}
    \item \textbf{MFB (Multimodal Factorized Bilinear Pooling):} We adopt the original factorized bilinear pooling strategy from \cite{yu2017multi} with matrix factorization rank set to $5$. 

    \item \textbf{LMF (Low-rank Multimodal Fusion):} Following the design in \cite{liu2018efficient}, we set the rank hyperparameter to $R=16$ in our experiments.

    \item \textbf{GCN (Graph Convolutional Network):} We implement one or two layers of GCNs, each followed by ReLU activation. The fused node representations are mean-pooled before going to the decoder.
    
    \item \textbf{Quantum Fusion Layer}: For the parameterized quantum circuits used in QFL, we adopt the hardware-efficient ansatz proposed by \cite{kandala2017hardware}. This design typically consists of repeated unit layers comprising parameterized single-qubit rotations interleaved with entangling two-qubit gates. In our implementation, each circuit $\mathbf{U}(\boldsymbol{\theta})$ consists of 5 such layers, where each layer applies a fully entangled block followed by rotation gates along the $x$-, $y$-, and $z$-axes. An illustration of the unit layer is provided in Figure~\ref{fig:pqcs}.
    \begin{figure}[!hbt]
        \centering
        \includegraphics[width=0.5\linewidth]{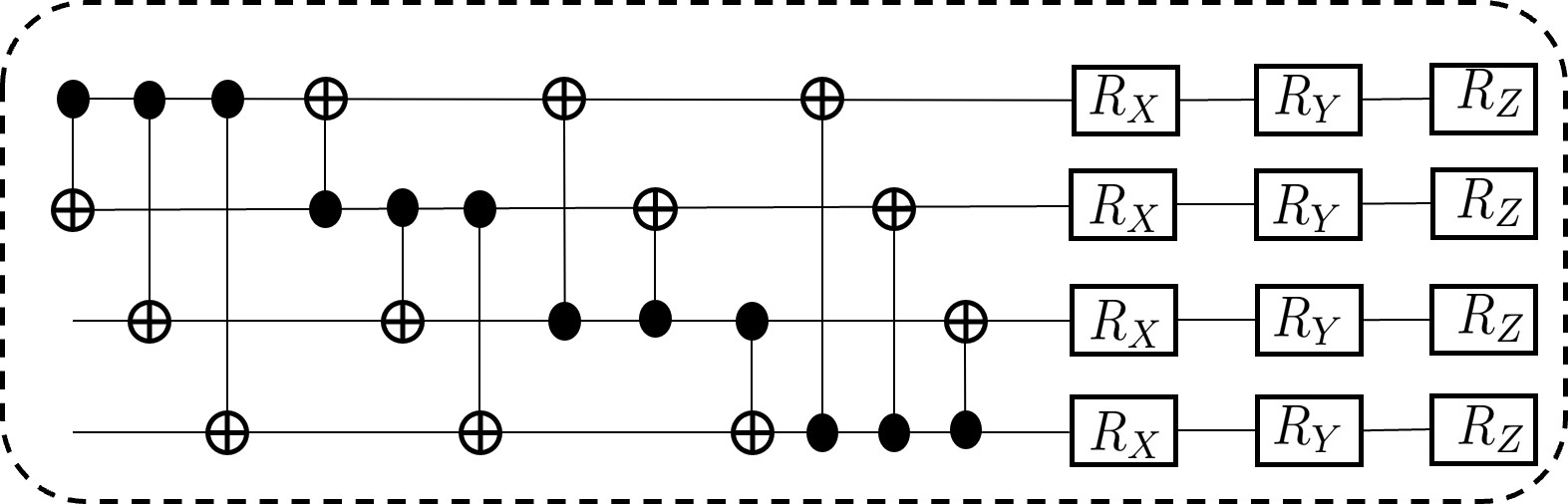}
        \caption[]{\textbf{Unit Layer of Parameterized Quantum Circuit used in our experiments}. $\mathbf{U}(\boldsymbol{\theta})$ consists of $5$ unit layers}
        \label{fig:pqcs}
    \end{figure}
\end{itemize}

\paragraph{Decoder.} For the decoder layer, we employ a simple fully-connected layer mapping the post-fusion output to the number of classes, depending on the dataset.

\subsection{Training Procedure}
All models are trained using the Adam optimizer with a learning rate of $10^{-3}$. We use a batch size of 32 for both the Multimodal Entailment and PTB-XL datasets, and a reduced batch size of 8 for Traffic-LA to accommodate memory constraints arising from the high modality count. Training proceeds for up to 100 epochs with early stopping, using a patience of 5 epochs based on validation performance. To address class imbalance in the Multimodal Entailment task, we employ the focal loss function, while standard cross-entropy loss is used for PTB-XL and Traffic-LA. All models, including baselines and QFL, are trained end-to-end under the same optimization settings to ensure fair comparison. The Quantum Fusion Layer (QFL) is implemented using TensorFlow Quantum, and all quantum simulations are performed using Cirq backend execution.

\end{appendix}

\end{document}